\documentclass[twoside,a4paper,11pt]{proceedings}
\usepackage{graphicx}
\usepackage{hyperref}
\usepackage{movie15}
\usepackage{natbib}
\usepackage{float}

\begin{document}
\pagenumbering{arabic}
\pagestyle{myheadings}
\thispagestyle{empty}
\vspace*{-1cm}
%{\flushleft\includegraphics[width=\textwidth,bb=58 650 590 680]{stamp.pdf}}
%{\flushleft\includegraphics[width=\textwidth,viewport=58 650 590 680]{stamp.pdf}}
{\flushleft\includegraphics[width=3cm,viewport=0 -30 200 -20]{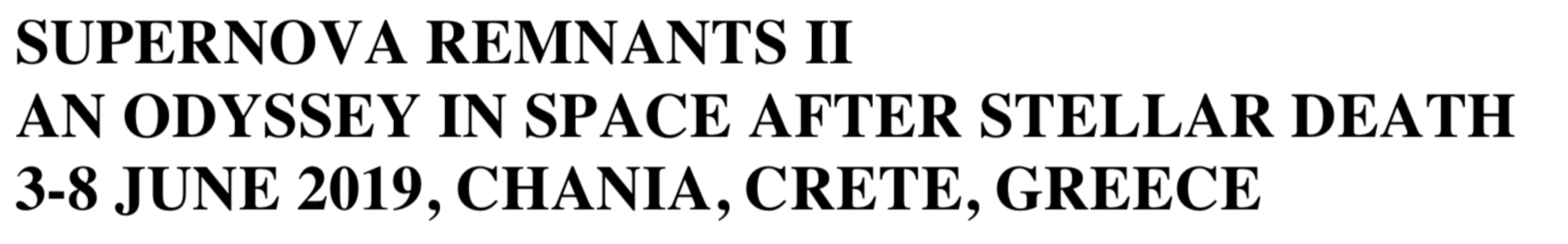}}
\vspace*{0.2cm}
\begin{flushleft}
{\bf {\LARGE
VRO 42.05.01: A supernova remnant resulted
by a supersonic, mass losing progenitor star}\\
\vspace*{1cm}
%%% Include here the LIST OF AUTHORS.
%%% Note that the last author has to be preceeded by an AND.
A. Chiotellis$^1$,
P. Boumis$^1$,
 S. Derlopa$^{1,2}$,
and W. Steffen$^3$
%,
%
% Do not delete next few lines
}\\
\vspace*{0.5cm}
%
%%% AFFILIATIONS LIST.
%%% and the AFFILIATIONS LIST. Note that one affiliation per line.
%%% Add as many affiliations as necessary. 
$^{1}$
Institute for Astronomy, Astrophysics, Space Applications
and Remote Sensing, \\
~~National Observatory of Athens,
15236 Penteli, Athens, Greece \\
$^{2}$
Department of Physics, University of Athens, Greece\\
$^3$  Institute of Astronomy, National Autonomous University of Mexico, Mexico

%
% Do not delete next few lines
\end{flushleft}
% Headings
\markboth{
%%% Type the SHORT version of the paper t
VRO 42.05.01: A SNR resulted by a supersonic mass losing progenitor star
}{
%%%  First Author \& Second Author   OR   First-author et al. 
%%%  First Author \& Second Author   OR   First-author et al. if the author list contains three or more authors.
Chiotellis et al. (2019)}
\thispagestyle{empty}
\vspace*{0.4cm}
\begin{minipage}[l]{0.09\textwidth}
\ 
\end{minipage}
\begin{minipage}[r]{0.9\textwidth}
\vspace{1cm}
\section*{Abstract}{\small
%%% Type the ABSTRACT of your paper

We present a model for the Galactic supernova remnant (SNR) VRO 42.05.01, suggesting that its intriguing morphology can be explained by a progenitor model of a supersonically moving, mass losing  star. The mass outflows of the progenitor star were in the form of an asymmetric stellar wind focused on the equatorial plane of the star.  The systemic motion of the parent star in combination with its asymmetric outflows excavated an extended wind bubble that revealed a similar structure to this of VRO 42.05.01. Currently the SNR is interacting with the wind bubble and it is dominantly shaped by it. Employing 2D hydrodynamic simulations we model VRO 42.05.01 under the framework of this model and we reproduce its overall morphological properties. We discuss the variations of our progenitor model in light of the current observational uncertainties. 

\normalsize}
\end{minipage}
%%% BODY of the paper

\section{Introduction}

\begin{figure}[h!]
\begin{center}
 \includegraphics[scale=0.22]{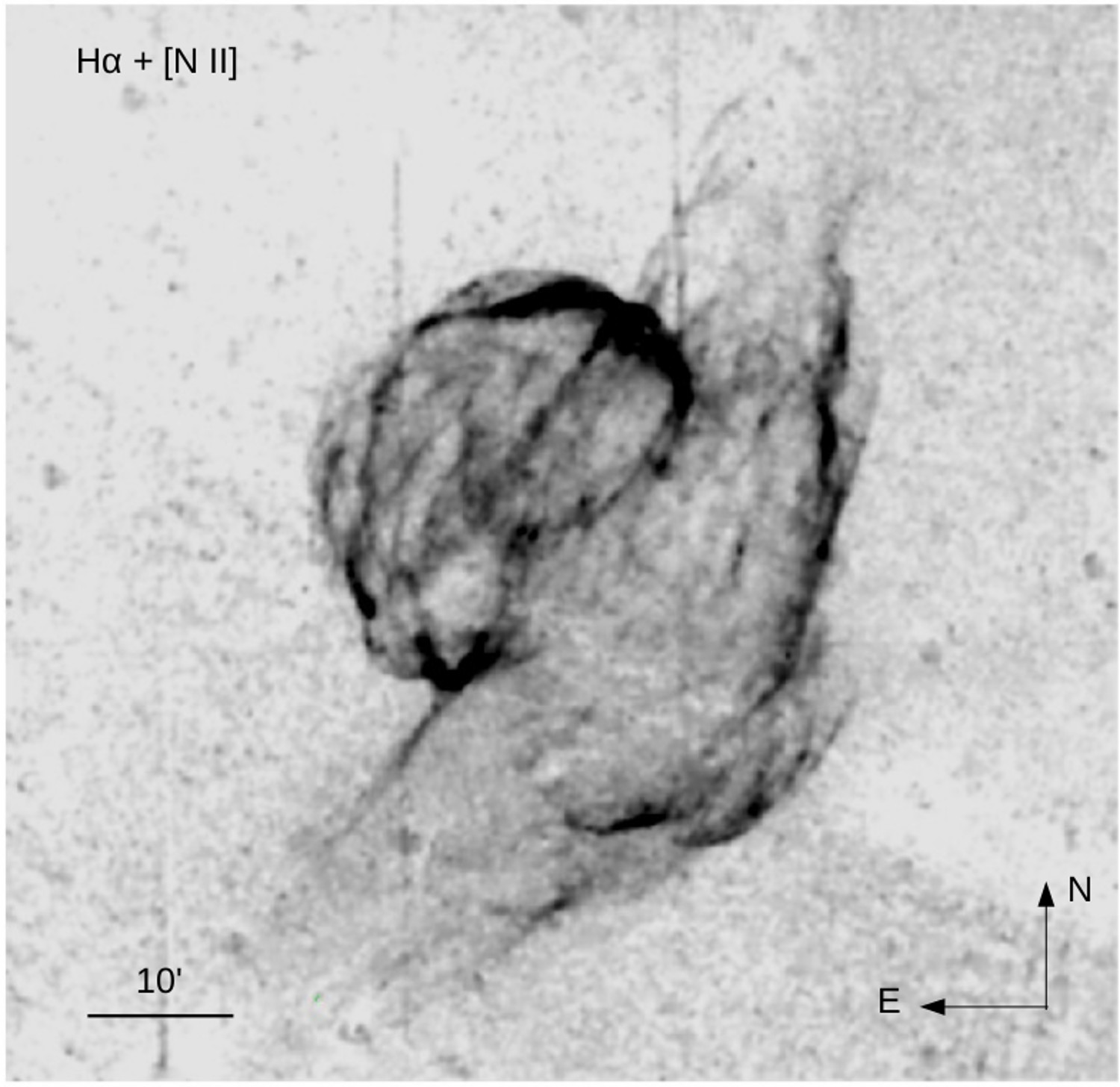}
 \caption{The $\rm H{\alpha} + [N~ {\sc II}]$  wide-field image of VRO 42.05.01 \citep{BOUMIS_SNRII_2019}}
  \label{fig:VRO}
  \end{center}
  \end{figure}

VRO 42.05.01 (hereafter VRO) is an evolved, optically bright Galactic SNR. The remnant reveals a very intriguing  morphology consisted of two major components: a$ ~30'$ diameter semicircular shell (a.k.a. `the shell') and a much larger bow shaped (almost triangular) shell (a.k.a. `the wing', see Fig. \ref{fig:VRO}). VRO's morphology indicates  that the remnant during its evolution encountered a non-homogeneous ambient medium and/or it resulted by an asymmetric supernova explosion.

Several models have been presented in the literature aiming to explain the peculiar morphology of VRO. These models consider that  VRO occurred close to the contact discontinuity of two interstellar media with different properties. According to them, the portion of the remnant that is evolving in the denser medium forms the `shell', while the `wing' results by the SNR 's breakout into the more tenuous medium  \citep{Pineault1985,Landecker1989}.  \citet{Arias2019a}  added an extra ingredient to these models suggesting that the progenitor of VRO was a runaway star. The supersonic motion of the progenitor formed a Mach cone around the explosion center. Subsequently, the interaction of the  SNR  with the surrounding Mach cone formed the triangular-shaped `wing'. However, recent observations of VRO and its environment did not reveal any  physical proof  of an interaction between the remnant and a non-homogeneous surrounding medium \citep{Arias2019b}. This fact in combination with the lack of any detailed modeling of VRO  retains its origin largely unknown. 

In this work, we present a novel model of VRO, which suggests that the SNR resulted by the interaction of the SN ejecta with the wind bubble formed by a supersonically moving, mass losing progenitor star. The  mass losses of the progenitor were in the form  of an asymmetric stellar wind,  focused on the equatorial plane of the star.  We demonstrate performing hydrodynamic simulations that this model can account for the distinctive morphology of the SNR.

\section{Methodology}\label{Sect:Method}
We employ the hydrodynamic code AMRVAC \citep{Keppens03} to simulate the wind bubble around the progenitor system and the subsequent evolution of the SNR. We perform the simulation in two steps. First we introduce the stellar wind in the axis origin of the grid in the form of a continuous inflow. The anisotropy of the wind is described by the following equations: 

\begin{equation} \label{eq:u_theta}
u_{w}(\theta)= u_{w, p} \left[1 - \alpha \mid \sin \theta \ \mid ^{k}  \right]~;~~ \rho(\theta)= \frac {\dot{M}_{p} \left[1 - \beta \mid \sin \theta \ \mid ^{k}  \right]^{-1}} {4 \pi r^2 u_{w}(\theta)}
\end{equation}\\
where $\rho(\theta)$,   $u_{w}(\theta)$ the wind density profile and the wind terminal velocity at the polar angle $\theta$, respectively.  $u_{w, p}$  is the wind terminal velocity and $\dot{M}_{p}$  the mass loss rate at the poles of the system ($\theta=0^{\rm o}$).  The $\alpha, \beta$ and $k$ are constants that determine the ratio of the polar velocity and density over the equatorial ones ($u_p/u_{eq}$ and $\dot{M}_{p}/\dot{M}_{eq}$, respectively) and the wind confinement at the equatorial plane.  Finally to simulate the supersonic motion of the progenitor star, woking on the star's rest frame, we let isotropic, homogenous gas to enter the grid antiparaller with the y-axis. After the formation of the wind bubble, we introduce in the inner boundary of the grid the SN ejecta with energy Eej  and mass Mej  and we let the SNR to evolve and interact with the ambient medium. The SN density profile is described by a power  law with index  n=9 while the velocity profile follows the Hubble expansion ($u \propto r_{SN}$).

\subsection{Wind bubble evolution}\label{Sect:windbubble}

Given the large size of VRO ($R_{Shell}= 15-30$ pc, for  VRO's  distance of $d=4.5 \pm 1.5$~kpc; \citealt{Landecker1989}), a powerful stellar wind is required in order to excavate an extended  bubble around the explosion center. The wind mechanical luminosity required by our models is in the order of $L= \frac{1}{2} \dot{M} u^2 \sim 10^{38}~ \rm erg~s^{-1}$. Such a high mechanical luminosity is best aligned to  stellar winds that emanate from Wolf-Rayet stars. Adopting this scenario and following the predictions of stellar evolution theory, we first simulate the formation of a wind bubble shaped by the wind of a Red Supergiant (RSG; i.e. the previous stellar evolutionary state before the Wolf-Rayet phase).   For the purposes of this simulation we set: $\dot{M}_{RSG}= 10^{-4}~\rm M_{\odot}~yr^{-1}$ and $u_{w,RSG}= 75~\rm km~s^{-1}$  while we let the wind to flow for $\tau_{RSG}= 4 \times 10^4~ \rm yr$.  Subsequently,  we impose the anisotropic Wolf-Rayet wind with properties: $\dot{M}_{WR}= 3 \times 10^{-5}~\rm M_{\odot}~yr^{-1}$,  $u_{w,WR}= 1000~\rm km~s^{-1}$ and $\alpha= 0.6$, $\beta= 0.95$ and $k =2$. Finally, the  interstellar medium properties we use are: $n_{ism}= 0.5~\rm cm^{-3}$ and $T_{ism}= 1000~\rm K$ while the systemic motion of the progenitor star is $u_*= 60~\rm km~s^{-1}$.

\begin{figure}[h!]
\begin{center}
 \includegraphics[scale=0.33]{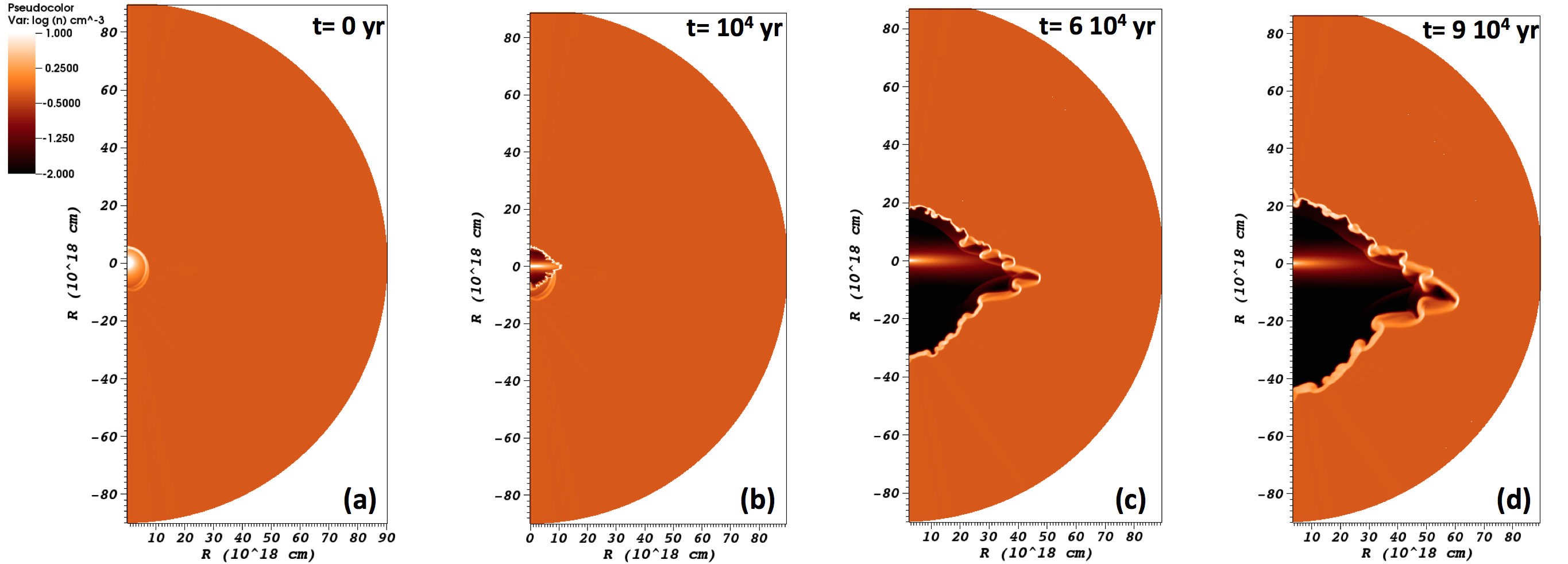}
 \caption{The 2D density contours of the wind bubble formed by the RSG wind (a) and of the  wind bubble evolution during the Wolf-Rayet progenitor's phase  (b, c, and d). }
  \label{fig:windbubble}
  \end{center}
  \end{figure}

The evolution of the wind bubble is illustrated in Fig. \ref{fig:windbubble}. The systemic motion of the progenitor star in combination with its asymmetric outflows excavate an extended wind bubble that reveals a similar morphological structure to this of VRO, consisting of the wing and shell component.

  \subsection{Supernova Remnant evolution}\label{Sect:SNR}

\begin{figure}[h!]
\begin{center}
 \includegraphics[scale=0.33]{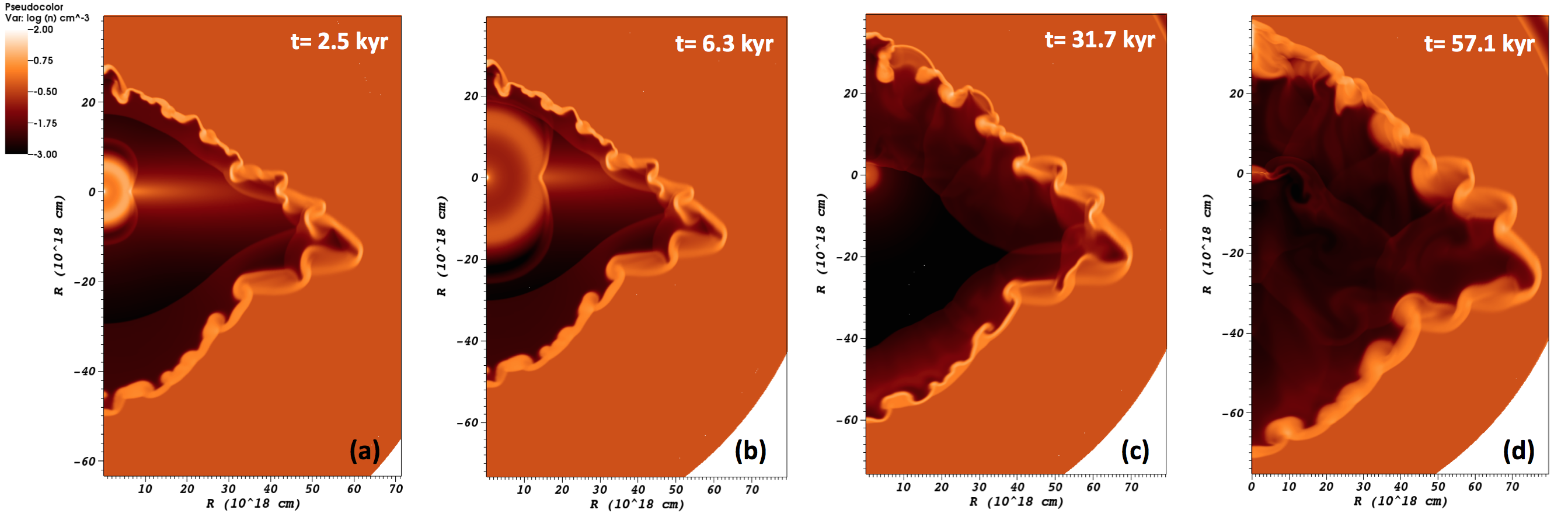}
 \caption{The 2D density contours of the SNR evolution within the wind bubble of Fig. \ref{fig:windbubble}d.}
  \label{fig:SNR}
  \end{center}
  \end{figure}

Subsequently, we introduce in the center of the circumstellar medium (CSM) of Fig. \ref{fig:windbubble}d the SN ejecta with energy $2 \times 10^{50}$ erg and mass $6~\rm M_{\odot} $ and we let the SNR to evolve.  The resulting SNR evolution is depicted in Fig. \ref{fig:SNR}. Initially, the SNR is within the wind bubble and it reveals a bi-lobal shape due to the CSM density enhancement at the equatorial plane of the system (Fig. \ref{fig:SNR}a). This bi-lobal shape is frequently met in mature SNRs \citep[e.g. G~65.3+5.7~;][]{Boumis2004}. Six thousand years after the explosion, the SNR starts to collide with the shell of the wind bubble and it is substantially decelerated (Fig. \ref{fig:SNR}b). The collision starts in the region of the wind bubble's  stagnation point  but progressively the SNR encounters the rest portions of the bubble. About thirty thousands years after the explosion, the SNR has interacted with the whole circumstellar structure and the remnant's morphology is dominantly shaped by it. Finally, 57 thousand years after the SN, the SNR reveals a morphology that represents the current state of VRO.

\section{Discussion}
\label{Sect: Concl}

We have presented evidence that the intriguing morphology of VRO can naturally be explained by the stellar wind properties of its progenitor star without the necessity of adopting peculiar ambient medium properties and  discontinuities. We argue that the observed characteristics of VRO can be reproduced by a supersonically moving progenitor star from which strong anisotropic stellar winds, enhanced at the equatorial plane, were emanating. Such anisotropic outflows can result by fast rotating stars and/or by close binary stellar systems.  The SNR simulations extracted by our 2D hydrodynamic modeling, show good agreement with the observed morphological properties of VRO. The  stellar wind properties of our  model are  most suitable to a Wolf - Rayet progenitor star. However,  the wind properties are determined by the natural size of VRO, which in turn is determined by the distance of the object. Given that VRO distance is not well defined \citep{Landecker1989, Arias2019b}, the properties of the anisotropic wind that shaped the cavity around VRO, and thus the nature of the progenitor star, are up to date,  not well constrained. A sequence of papers relevant to VRO  will be published soon (VRO's optical observations; \citet{BOUMIS_SNRII_2019} and VRO's 3D morphokinematical modeling; \citealt{Derlopa2019})  that are expected to limit the parameter space of the remnant's physical quantities  and consequently provide vital insights on the unknown nature of VRO 42.05.01.

% Do not delete the next line
\small  % Do not delete
\section*{Acknowledgments}   % Do not delete if you declare acknowledgments
We thank Rony Keppens for providing us with the AMRVAC code. A.C. gratefully acknowledges Th.~Parizas for the fruitful discussions and motivation. A.C. and S.D. acknowledge the support of this work by the PROTEAS II project (MIS 5002515), which is implemented under the ``Reinforcement of the Research and Innovation Infrastructure" action, funded by the ``Competitiveness, Entrepreneurship and Innovation" operational programme (NSRF 2014-2020) and co-financed by Greece and the European Union (European Regional Development Fund). S.D acknowledges the Operational Programme ``Human Resources Development, Education and Lifelong Learning'' in the context of the project ``Strengthening Human Resources Research Potential via Doctorate Research'' (MIS-5000432), implemented by the State Scholarships Foundation (IKY) and co-financed by Greece and the European Union (European Social Fund- ESF). 
%%% BIBLIOGRAPHY
\bibliographystyle{aj}
\small
\bibliography{VRO_hydro_Chania}

\end{document}